\documentclass[twocolumn,aps,showpacs,amsmath,amssymb]{revtex4}
\usepackage{graphicx}
\usepackage{psfrag}
\usepackage{epsfig}
\usepackage{rotating}
\usepackage{color}

\begin{document}

\title{The Role of Second Trials in Cascades of Information over Networks}
\author{C. de Kerchove$^{1}$\email{c.dekerchove@uclouvain.be}}
\author{G. Krings$^{1}$\email{gautier.krings@uclouvain.be}}
\author{R. Lambiotte$^{1,2}$\email{r.lambiotte@imperial.ac.uk}}
\author{P. Van Dooren$^{1}$},
\author{V.D. Blondel$^{1}$}
\affiliation{
  $^1$ INMA, Universit\'e catholique de Louvain,  4 avenue Georges Lemaitre,
B-1348 Louvain-la-Neuve, Belgium \\
$^{2}$ Institute for Mathematical Sciences, Imperial College London, 53 Prince's Gate, South Kensington campus, SW7 2PG, UK}

\date{\today}

\begin{abstract}
We study the propagation of information in social networks. To do
so, we focus on a cascade model where nodes are infected with {
probability $p_1$ after their first contact with the information
and with probability $p_2$ at all subsequent contacts.} The
diffusion starts from one random node and leads to a cascade of
infection.  It is shown that first and {subsequent} trials play
different roles in the propagation and that the size of the
cascade depends in a non-trivial way on $p_1$, $p_2$ and on the
network structure. Second trials are shown to amplify the
propagation in dense parts of the network while first trials are
{dominant for the exploration of} new parts of the network and
launching new seeds of infection.
\end{abstract}

\maketitle

% main text
\section{Introduction}
%\label{}

The propagation of information and new ideas has long been a
fundamental question in the social sciences. Propagation may be
driven by exogenous causes, when people are informed in a {\em
mean-field} way by an external source, e.g. television, but also
by endogenous mechanisms, when a few early adopters may influence
their friends, who may in turn influence their own friends and
possibly lead to a cascade of influence {\cite{sornette}}. This
self-organizing process, which reminds of the dynamics of an
epidemic, is usually called the word-of-mouth phenomenon. It has
attracted more and more attention in the last few years due to the
emergence of the internet and of online social networks, which
have led to more decentralized media of communication. A typical
example is the blogosphere, where blogs are written and read by
web users and where debates/discussions may take place among the
bloggers. As of today, the blogosphere is extremely influential in
the adoption or rejection of products but also in politics, as
more and more citizens {voice} their opinions and mobilize
community efforts around their candidates. From a practical point
of view, the emergence of these participative media has changed
the way elections take place, by allowing politicians to reach new
audiences, raise money, communicate to voters and even consider
all of them as a gigantic think tank \cite{obama}, and also to
open new ways to promote commercial products via recommendation
networks or viral marketing methods. It is therefore interesting
to better understand how such information cascades take place in
social networks
\cite{granovetter1978tmc,kempe2003msi,goldenberg2001tnc,gruhl2004idt,leskovec2006pir,
watts2007ina}.

A good description of the word-of-mouth phenomenon requires two
elements: {a model of propagation and a network structure}. The
model of propagation defines the way information (e.g. a marketing
campaign for a specific product, an information) flows between
acquaintances. One of the most common models of propagation is the
\textit{Independent Cascade Model}
(ICM)\cite{kempe2003msi,goldenberg2001tnc}, where one starts from
an initial set of infected nodes. When a new node becomes
infected, it tries one single time to infect each of its neighbors
with independent probability $p$. The process stops when no new
node has been infected. The size of the information cascade is
given by the number of infected nodes and one says that an
epidemic outbreak (keeping in mind that the models described in
this paper apply only to information diffusion, not to the
epidemical spread of diseases) takes place when the fraction of
people who are infected does not vanish {as the network size
increases}. It is straightforward to show that ICM is equivalent
to the epidemiological SIR model, where nodes are divided in three
classes, i.e. susceptible/infectious/removed
\cite{PhysRevE.66.016128}, and where infectious nodes infect their
neighbors with rate $p$ and are removed with rate $1$. It is also
possible to view ICM as a bond percolation problem, the final
number of infected nodes being the sum of the sizes of the
connected components the initial nodes belong to. {Second,} this
viral process has to be applied on a realistic social network,
where each node defines a member of the society and edges are
drawn between acquaintances. For a long time the design of these
social networks was purely theoretical and real social networks
were generally limited in size, but the advent of the Internet and
of cheap computer power now allows to study social networks
composed of millions of individuals and to characterize the
statistical properties of their topology. For instance, it has
been shown that social networks typically exhibit the small-world
property \cite{sw}, heavy-tailed degree distributions \cite{bara},
assortative mixing \cite{newman00}, modular structure
\cite{modular}, etc. An important challenge is therefore to
understand how the topology of the social network affects the
propagation of information but also to find statistical indicators
for the most influential nodes in the network \cite{kempe2003msi,
kimura2006tmi, blondel2006slg,blondel2007}.

The ICM is a direct implementation of an epidemiological model in
a social context. There are, however, drastic differences between
the propagation of a virus and the propagation of ideas. Indeed,
recent experiments have shown that the memory of the individuals
may play a dominant role in the latter case. For instance, in the
case of recommendation networks, the probability that people buy
an item depends in a non-trivial way on the number of times they
received a recommendation for this item \cite{leskovec2007dvm}. In
the case of online social networks, it was also shown that the
probability to join a community depends on the number of your
friends in that community \cite{bhkl06}. In general, empirical
studies show that the probability of getting infected increases
with the number of contacts $k$ and saturates for large values of
$k$. Several models have been introduced in order to take into
account this property, such as general ICM, threshold and cascade
models
\cite{joo2004sat,watts2002smg,dodds2004ubg,kleinberg:cbn,klimek}
or generalized voter models \cite{galam,lambiotte}. The way such
dynamics is affected by the network topology is, however, still
poorly understood \cite{ganesh}, {even though some studies focus
on specific topologies~\cite{gleeson1,gleeson2,centola,centola2}}.
The goal of this paper is to bridge this gap by focusing on a
generalization of ICM which includes in the simplest way a
dependence on the number of contacts. The model is applied on
small-world networks in order to highlight the importance of the network randomness. As a first step, we focus on simplified cases where the network is directed, which allows us to obtain an analytical description of the propagation. It is shown that the birth of large
cascades of information is strongly influenced by the network
topology and that first and subsequent trials play very different
roles in the propagation. Computer simulations are also performed on directed and on more realistic undirected networks, and confirm the above observations.

\begin{figure}[h]
\centering
\includegraphics[width=8cm]{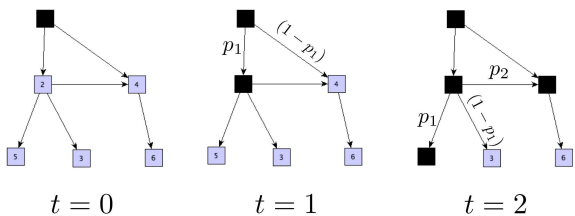}
\caption{Illustration of the generalized ICM. Infected nodes
contact their neighbours only once. These neighbours get infected
with probability $p_1$ if it is the first time they are contacted
(and therefore remain uninfected with probability $1-p_1$) and $p_2$ otherwise. The presence of triangles and, by extension
of local structures, is crucial for second and subsequent trials
to be frequent.}\label{model}
\end{figure}

\section{Properties of the model}

Our generalization of ICM is defined as follows. The network is
composed of $N$ nodes and one node is initially infected. Each
time a new node is infected, it contacts all of its neighbours,
and they each get infected with a probability $p_1$ if it is the
first time they are contacted and with a probability $p_2$ {for
all subsequent contacts}. The dynamics stops when no new node is
infected. The classical ICM is therefore recovered when $p_1=p_2$.
Since the ICM and SIR model are equivalent, one can also interpret
the generalized ICM as an extension of the SIR model. The
dependence in the number of contacts leads to a new class of
nodes, namely \textit{contacted} nodes, which have already been
unsuccessfully attacked by infectious nodes. In that framework,
the probability of a susceptible node to be infected by a
neighboring infectious node is $p_1$ while it becomes contacted
with probability $1-p_1$. When a contacted node is attacked by an
infectious node, its probability to become infected is $p_2$.
Finally, an infectious node becomes removed once it has attacked
each of its neighbors. The model can also be related to threshold
models \cite{granovetter1978tmc,kempe2003msi} where each node
receives a random threshold generated following a given
distribution. A node becomes infected when the number of infected
neighbors exceeds his threshold. The probability of having a
threshold of value 1 is the probability of being infected at the
first trial, in our case $p_1$. In this way, one can generate for
every couple ($p_1$,$p_2$) the thresholds of the equivalent
threshold model with the following expressions
\begin{eqnarray}\label{eq:pToTheta}
P(\theta = 1) & = & p_1\\
P(\theta = k) & = & (1-p_1)(1-p_2)^{k-2}p_2\qquad\forall\phantom{.}k\geq2.
\end{eqnarray}
It is also interesting to note that our model may be related to
percolation. The case $p_1=p_2$ is well-known to be equivalent to
bond percolation but the case $p_2=0$ can also be seen as a node
percolation problem. Indeed, in that case, each neighbour of an
infected node is infected with a probability $p_1$ only if it is
the first time it is in contact with the information. The total
number of infected nodes may therefore be obtained by removing
nodes from the network with a probability $1-p_1$ and by looking
at the size of the connected components. For general values of
$p_1$ and $p_2$, however, the system is much more complicated and
the probabilities of infection are not straightforward to compute.

\section{Random Networks}

In this paper, we are interested in the conditions for a large
cascade to emerge. We therefore look for the critical couple
($p_{1_c}$,$p_{2_c}$) such that a random node infects a non
vanishing fraction of the network for any couple ($p_{1}$,$p_{2}$)
$\geq$ ($p_{1_c}$,$p_{2_c}$) where the inequalities are
componentwise. This couple determines the epidemic threshold of
this network. Let us first focus on a directed random
Erd{\"o}s-Renyi network, composed of $N$ nodes and where the
probability to have a link between two randomly selected nodes is
$p_{er}$. As we will show, the proportion of second attacks
vanishes when $N$ tends to infinity when one is below the epidemic threshold. Therefore the threshold in
such topology hardly depends on $p_2$ and we recover the same threshold
as for the ICM model. Even though this result was predictable, the
probability $p_2$ still plays a role when the size of the network
is finite. Let $S(t)$, $C(t)$, $I(t)$ and $R(t)$ be the number of
susceptible, contacted, infectious and removed nodes respectively
at time $t$. By using a mean-field approximation, one obtains the
number of links between different types of nodes. For instance,
the number of links going from infectious nodes to susceptible
nodes is $S(t)I(t)p_{er}$, which also represents the number of
attacks at time $t$ from infectious nodes on susceptible nodes.
The average number of susceptible nodes that become infected at
time $t$ is therefore given by $S(t)I(t)p_{er}p_1$. Similar
calculations lead to the set of equations
\begin{equation}\label{eq:scir_dynamics}
\left\{\begin{array}{ccl}
\dot{s} & = & -s\,i\,d\\
\dot{c} & = & s\,i\,d\,(1-p_1)-c\,i\,d\,p_2\\
\dot{i} & = & -i + s\,i\,d\,p_1 + c\,i\,d\,p_2\\
\dot{r} & = & i
\end{array}
\right.
\end{equation}
for the densities $s$, $c$, $i$ and $r$, where $s=\frac{S}{N}$,
$c=\frac{C}{N}$, $i=\frac{I}{N}$, $r=\frac{R}{N}$, and where $d =
N \, p_{er}$ is the average degree of the network.

The epidemic threshold is found by linearizing {this nonlinear
dynamical system} around the stationary solution $\boldsymbol{x}_0
= (1,0,0,0)$ where all nodes are susceptible, and looking at the
eigenvalues of the linearized matrix. {The behaviour of the system
is then essentially governed by the linearized equation $\dot{i} =
i(-1+d\,p_1)$, which implies that} $\boldsymbol{x}_0$ is stable if
$d\,p_1 < 1$ and therefore that the infection will not reach a non
vanishing fraction of the network in that case. This result, which
is well known in percolation theory when $p_2=p_1$, also shows
that the epidemic threshold does not depend on the parameter
$p_2$. This may be understood by noting that second and subsequent
trials are statistically relevant only when a finite fraction of
nodes have been infected, which implies that the epidemic
threshold may be evaluated without taking them into account.
This also implies that for a non-vanishing initial fraction of contacted
nodes, we then have a dependency on $p_2$ and the threshold will
change accordingly. Above the epidemic threshold, the system of
equations (\ref{eq:scir_dynamics}) ceases to be valid because it
does not incorporate multiple attacks (i.e. several edges
attacking a node at the same time), thereby leading to an
overestimation of the number of infections. In that case, we have
therefore performed computer simulations of the model which show
that the total fraction of nodes $r(\infty)$ having been infected
increases with $p_2$, as expected. This becomes even more
obvious for $N$ decreasing. Finally, when $N$ is relatively small,
the proportion of second attacks is no more negligible and the
threshold varies with $p_1$ and $p_2$.

\section{Directed Small-World network}

\begin{figure}[h]
\centering
\includegraphics[width=9cm]{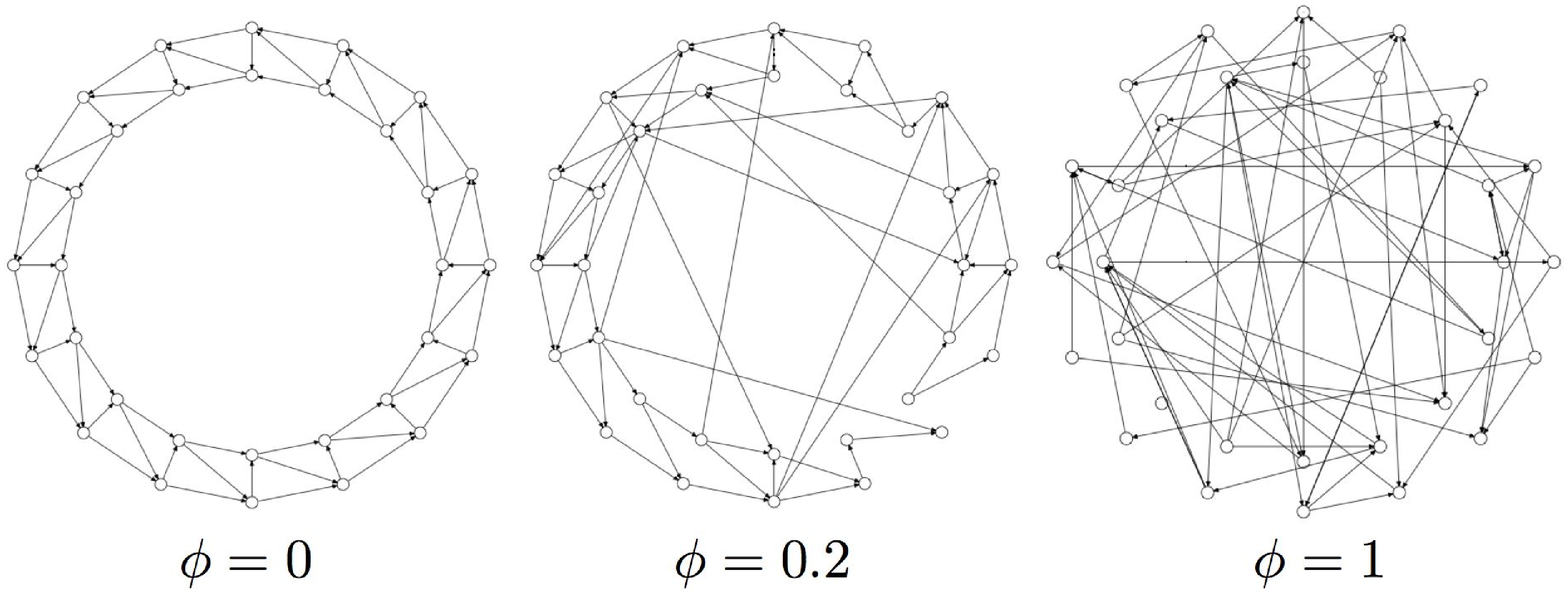}
\caption{For different values of $\phi$, the topology is a regular lattice ($\phi=0$),
a small world network ($\phi=.2$) or a random network ($\phi=1$).}
\label{lattice}
\end{figure}

In order to highlight the role played by the network topology, we
have applied the model on a directed version of the well-known
Watts-Strogatz model for small-world networks \cite{sw}. The
main reason for looking at this directed version rests in the
equations of propagation that becomes tractable. However the
simulations show that both cases, directed and undirected, exhibit
similar couple of thresholds. The directed version is built from
a directed one-dimensional lattice of $N$ sites, with periodic
boundary conditions, i.e., a ring, each vertex $k$ pointing to 2
neighbors $k+1$, $k+2$, see Fig.~\ref{lattice}. With probability
$\phi$, these ``regular'' links are removed and replaced by random
links. This network therefore exhibits an interplay between order
and randomness. By increasing the parameter $\phi$, one increases
the randomness of the topology and one recovers a random network
when $\phi=1$.

\begin{figure}
\includegraphics[width=8cm]{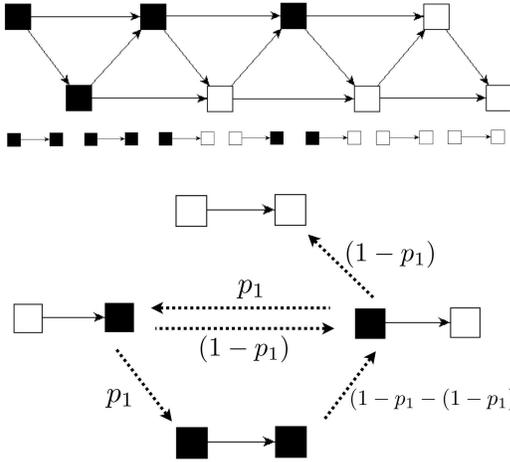}
\caption{On a regular lattice, the states of the nodes $k$ and $k+1$, denoted by $i$ and $j$ respectively, fully determine the state of node $k+2$. The dynamics is therefore specified by the succession of states $(i,j)$. The dynamics ends when two successive zeros, i.e. a state $(0,0)$, take place.}\label{fig:markov_chain}
\end{figure}

It is instructive to first consider the case of a regular lattice,
i.e., $\phi=0$. In that case, the information propagates in the
system in an ordered way and the state of each site $k$ is only
influenced by the sites $k-2$ and $k-1$. {For this reason one
does not need to store separately the state ``contacted''
anymore.} Let $n_{ij;k}$, with $i,j \in \{0,1\}$ be the
probability that node $k$ is $i$ and node $k+1$ is $j$, with the
correspondence: $1=$ infectious, $0=$ not infectious. By
definition, $\sum_{i,j} n_{ij;k}=1$ for any $k$. Let us assume
that one starts the propagation at node $k=1$, so that
$n_{01;0}=1$. Then, it is straightforward to show that the
quantities $n_{ij;k}=1$ satisfy the recurrence

\begin{eqnarray}\label{matrix}
n_{11;k+1} &=& (p_1 + (1-p_1) p_2) n_{11;k} + p_1
n_{01;k} \cr n_{01;k+1} &=&  p_1 n_{10;k} \cr n_{10;k+1} &=& (1 -
p_1 - (1-p_1) p_2) n_{11;k} + (1-p_1) n_{0,1}\cr
\end{eqnarray}
while the probability that the dynamics ends grows monotonically
like
\begin{eqnarray}
n_{00;k+1} &=& n_{00;k} + (1-p_1) n_{10;k}.
\end{eqnarray}
This corresponds to the $4$ states Markov chain represented in
Fig.~\ref{fig:markov_chain}. {By definition, the expected number
of infected nodes is $N_{\infty}=\frac{1}{2}\:\sum_{i,j,,k}
n_{ij;k+1}$.}

The asymptotic number of infected nodes grows like the largest eigenvalue of the matrix associated with the linear system (\ref{matrix})
\begin{eqnarray}\label{eq:A}
\boldsymbol{A}= \left[
\begin{array}{ccc}
p_1+(1-p_1)p_2 & p_1 & 0\\
0 & 0 & p_1 \\
(1-p_1)(1-p_2) & (1-p_1) & 0
\end{array}
\right].
\end{eqnarray}
This largest eigenvalue is smaller than $1$ for any $p_1$, $p_2$,
except when $p_1=1$ or $p_2=1$, which implies that an epidemic
outbreak takes place only in these trivial cases. In contrast,
when $p_1$ and $p_2$ are different from $1$, only a finite number
of nodes gets asymptotically infected.  This is due to the
one-dimensionality of the topology, which implies that two nodes
at most may spread the infection at each step and that the
probability that no new node gets infected is different of zero
when $p_1\neq 1$ and $p_2\neq 1$. As expected, increasing values
of $p_1$ or $p_2$ increase the total number of infected nodes.
{The analytical expression for $N_\infty$ when $p_1,p_2\neq0$ is
given by
\begin{eqnarray}
N_{\infty}&=&\frac{1}{2}[\begin{array}{ccc}2&1&1\end{array}]\sum_{k=0}^{\infty}\boldsymbol{A}^k
[\begin{array}{ccc}0&1&0\end{array}]^T,\cr
&=&\frac{1}{2}
[\begin{array}{ccc}2&1&1\end{array}][\boldsymbol{I}-\boldsymbol{A}]^{-1}
[\begin{array}{ccc}0&1&0\end{array}]^T,\cr
&=& \frac{(1-p_1)(1-p_2)+p_1}{(1-p_1)^2(1-p_2)}.
\end{eqnarray}}

%It is in general not possible to write an analytical expression
%for $N_{\infty}$, except when $p_2=0$ and the problem is
%equivalent to node percolation. In that case, let us consider the
%probability $n_i$ that $i$ nodes are infected before the infection
%stops. It is straightforward to show that
%\begin{eqnarray}
%n_i = (1-p)^2 p^{i-1} \sum_j \binom{j}{i-1}  (1-p_1)^j= (1-p_1)^2 p_1^{ i-1} (2-p)^{i-1},
%\end{eqnarray}
%so that the average number of infected nodes  is
%$N=\sum_i i n_i=1/(1-p_1)^2$.

Let us now focus on a topology where a fraction of the links is
displaced in a random way. In order to generalize the results of
the previous section, it is useful to label each node with its
position $k$ on the underlying one-dimensional lattice. By
construction, each node $k$ points to $k+1$ and $k+2$ when
$\phi=0$ but such links only exist with probability $1-\phi$ in
general. In a system where $\phi$ is sufficiently small and where
only a vanishing fraction of the nodes gets activated, one may
decouple the dynamics as follows \cite{moore}. The initial seed
may infect a segment of nodes which are contiguous on the
underlying lattice, thereby leading to $N_1(p_1,p_2)$ contiguous
infected nodes. This number may be evaluated by generalizing the
set of equations (\ref{matrix}) and taking into account the fact
that some links are missing. The associated matrix with this
linear system is {
\begin{eqnarray}\label{eq:A phi}
\boldsymbol{A}_\phi &=& (1-\phi)^2 \boldsymbol{A} +
2\phi(1-\phi)\left[
\begin{array}{ccc}
p_1 & \frac{p_1}{2} & 0\\
0 & 0 & \frac{p_1}{2} \\
1-p_1 & 1-\frac{p_1}{2} & 0
\end{array}
\right] \cr
&+& \phi^2 \left[
\begin{array}{ccc}
0&0& 0\\
0 & 0 & 0 \\
1 & 1 & 0
\end{array}
\right],
\end{eqnarray}}
where we consider three cases: no missing link, this occurs with
probability $(1-\phi)^2$, and then we recover the matrix
$\boldsymbol{A}$ in Eq.~\ref{eq:A}, secondly we have with
probability $2\phi(1-\phi)$ one missing link and the corresponding
transition matrix, {and finally we have with probability $\phi^2$
no link accompanied by a simple transition matrix. By using
similar arguments that for the regular lattice, one finds that the
average number of contiguously infected nodes is
\begin{widetext}
\begin{eqnarray}
\label{n1}
N_{1}(p_1,p_2) &=& \frac{1 - (1-p_1) p_2 (1-\phi)^2 - p_1 (1-\phi) \phi}{(1-p_1 (1-\phi))(1-p_2 (1-\phi)^2 - p_1 (1-\phi) (1- p_2 (1-\phi) + \phi))}.
\end{eqnarray}
\end{widetext}

This segment of $N_1(p_1,p_2)$ infected nodes may in turn infect $2 \phi p_1 N_1(p_1,p_2)$ distant nodes which will play the role of a new seed, each of them infecting a new segment of average size $N_1(p_1,p_2)$, etc. Below the epidemic threshold, only a vanishing proportion of nodes is infected and one may assume that the different segments do not overlap. The total number of infected links is therefore
\begin{eqnarray}
N_{\infty} = N_1(p_1,p_2) \sum_{i=0}^{\infty} (2 \phi p_1
N_1(p_1,p_2))^i
\end{eqnarray}
which converges to
\begin{eqnarray}
N_{\infty} = N_1(p_1,p_2)/(1-2 \phi p_1 N_1(p_1,p_2))
\end{eqnarray}
if
\begin{eqnarray}
\label{final} 2 \phi p_1 N_1(p_1,p_2)<1.
\end{eqnarray}

\begin{figure}[h]
\centering
\includegraphics[angle=-90,width=9cm]{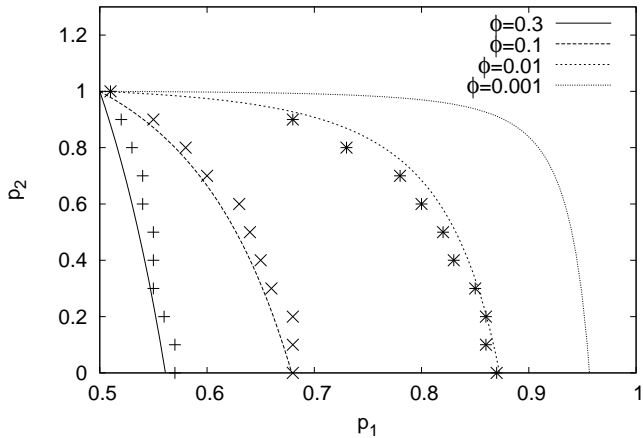}
\caption{Couples of thresholds for $\phi=[0.3 \:\: 0.1\:\: 0.01
\:\: 0.001]$. In the limit $\phi \rightarrow 1$ of a random
network, the critical line is vertical, $p_1=1/2$, and the
epidemic threshold is therefore independent on $p_2$. The
different signs represent experimental couples of threshold
obtained with a precision of $0.01$. The slight shifts with
respect to the theoretical curves come from the finite size of the
network ($10^4$ nodes), this effect increases for $\phi$ close to
$0$.}\label{trans_2}
\end{figure}

\begin{figure}[h]
\centering
\includegraphics[angle=-90,width=9cm]{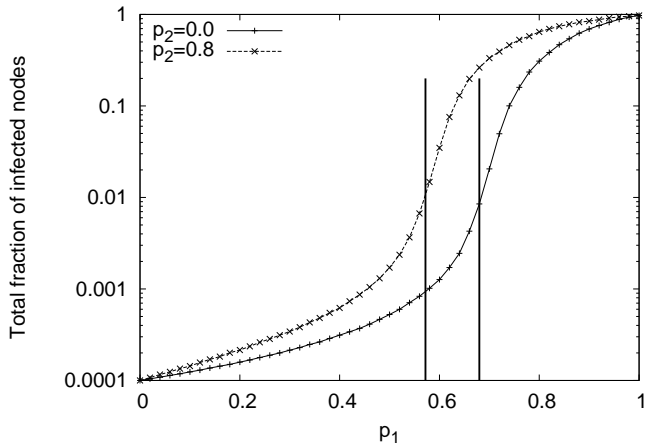}
\caption{Total fraction of infected nodes in log scale as a
function of $p_1$, for $p_2=0$ and $p_2=0.8$ respectively. The
network is composed of $10^4$ nodes {and $\phi=0.1$}. Vertical
lines correspond to the theoretical prediction $p_{1c}$ where
cascades occur.}\label{simu}
\end{figure}

The line $2 \phi p_1 N_1(p_1,p_2)=1$ therefore separates two
regimes, one in which the spreading dies out and another one in
which an infinite number of nodes is asymptotically infected. {By
using Eq.(\ref{n1}) and solving Eq.(\ref{final}), one finds an
analytical formula for the critical value
\begin{eqnarray}
p_{2c} = \frac{1 - (2 + \phi - \phi^2) p_1+ (1- \phi + \phi^2 -
\phi^3)p_1^2}{ (1 - \phi)^2(1-p_1)  (1 - p_1 - p_1 \phi)},
\end{eqnarray}
such that an epidemics takes place when $p_2>p_{2c}$ (see
Fig.~\ref{trans_2}).} It is interesting to note that the epidemic
threshold  depends both on $p_1$ and $p_2$ for general values of
$\phi$, but that these parameters are associated with different
mechanisms. The probability $p_2$ plays an important role in the
local propagation of the infection among neighbouring sites. The
probability $p_1$ also plays a role for such propagations but it
is also responsible for the infection of new distant seeds, a
process {that is crucial for exploring} several disconnected parts
of the network and that favours the emergence of an epidemic. One
observes from { \eqref{n1}} and \eqref{final} that $p_2$ is less
and less important as $\phi$ increases. In the limit $\phi
\rightarrow 1$ of a random network, the length of infected
segments $N_1(p_1,p_2)$ goes to 1, which implies that the epidemic
threshold is $p_1=1/2$, independently of $p_2$, as predicted in
our analysis of the Erd{\"o}s-R{\'enyi} network. It is also
interesting to note that the total number of infected nodes (11)
may decrease when $\phi$ is increased, which is in contradiction
with the usual belief that short-cuts promote the propagation
\cite{centola, centola2}.

We have checked the validity of (\ref{final}) by performing
computer simulations of the generalized ICM on a directed
small-world network with $N=10^4$ nodes and by averaging the
results over {$10^4$} realizations of the dynamics. As shown
in Fig.~(\ref{simu}), the critical threshold for a given
$p_2$ is evaluated by looking at the probability $p_1$ for which
the slope of $N_{\infty}$ is maximal when the Y-axis is in
log-scale. In Fig.~(\ref{trans_2}) these critical points are drawn
for $\phi=0.3$, $\phi=0.1$ and $\phi=0.01$. The case $\phi=0.001$
is not shown because of the very small number of short-cuts in
that case and therefore of the very large fluctuations from one
realization of the network to another one. The simulation results
show large fluctuations but are nonetheless in good agreement with
the theoretical predictions.

Finally, we have also studied numerically our model when it is applied
to an undirected small-world network made of $10^4$ nodes and with an average degree $4$. As expected (the mean degree is twice larger),
the frontiers are shifted to the left meaning that smaller
probabilities are sufficient to observe significant cascades in
the network (see
Fig.~\ref{simu2}). Qualitatively, however, the system behaves in the same way as in the directed case and the lines determining the epidemic threshold have similar shapes. Theoretically, when $\phi=1$, the network is random and the epidemic threshold should not depend on $p_2$, i.e. it is
 a vertical line. However, the finite size of the network implies that the
proportion of triangles does not vanish and therefore that second attacks may occur due to finite size effects. Consequently the experiments show a slight dependency on
$p_2$ and the frontier is not exactly vertical when $\phi=1$.
However we recover the threshold of the ICM model when
$p_1=p_2=0.25$. 

\begin{figure}[h!]
\centering
\includegraphics[angle=-90,width=9cm]{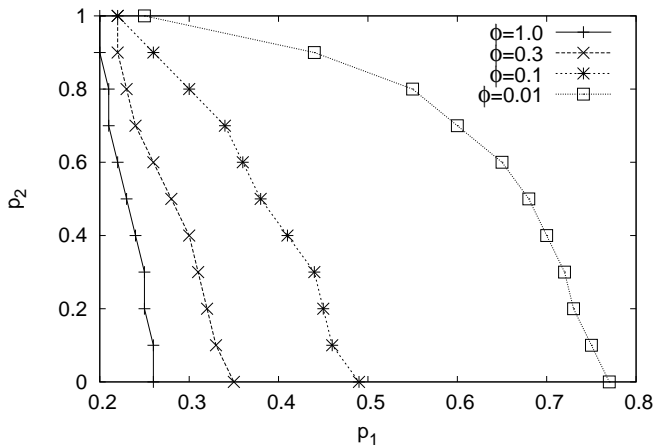}
\caption{Experimental couples of thresholds for $\phi=[0.3 \:\:
0.1\:\: 0.01]$ when the small-world is undirected and made of
$10^4$ nodes. The frontiers of transitions are shifted to the left
and they exhibit similar shape.}\label{simu2}
\end{figure}

\section{Conclusion}

In this paper, we have focused on a very simple model for the
cascade of information in social networks. The novelty of the
model consists in considering different probabilities for being
infected depending on the number of contacts with the information.
The model has been applied on a directed small-world network in
order to show how the randomness of the network topology affects
the propagation. It is shown that first and subsequent trials play
very different roles: first trials are primordial in order to
discover unexplored parts of the network and launch new seeds of
infection, while second and subsequent trials influence the
propagation in ordered parts of the network, where triangles (and
other dense motifs) are frequent. The epidemic threshold, which
determines the success of the cascade, depends in a non-trivial
way on these two mechanisms and on the randomness of the network
topology, but it is dominated by the success of first trials.

The importance of first trials should be put in perspective with
Granovetter's famous work on ``The Strength of Weak Ties''
\cite{granovetter2,onnela}, which states that  weak links keep the
network connected whereas strong links are mostly concentrated
within communities. In the context of information diffusion, our
model shows that the first trials play a similar cohesive role by
connecting different communities, while second and subsequent
trials accelerate the propagation inside the communities. This is
due to the fact that dense parts in the network make possible the
existence of several infected paths to each node, and therefore
increase the number of time one node is contacted. In the extreme
scenario of a $k$ clique, for instance, where $k$ nodes are fully
connected, after the first step, all further steps will be
considered as second trials.

To conclude, our model is motivated by recent experiments which
have shown that an accumulation of contacts favours the
propagation of information and that, in particular, second and
subsequent trials are more successful than first trials.
Interestingly, our model also reproduces the fact that locally
dense subnetworks accelerate the propagation
\cite{mcadam1,mcadam2}, a property which has been observed for the
adoption of new services among users of a mobile phone networks
\cite{prieur} and which is not reproduced by the original ICM.

\medskip
\noindent
{\bf Acknowledgements}

{This work has been supported by the Concerted Research Action
(ARC) ``Large Graphs and Networks'' from the ``Direction de la
recherche scientifique - Communaut\'e fran\c{c}aise de
Belgique.'', by the EU HYCON Network of Excellence (contract
number FP6-IST-511368), and by the Belgian Programme on
Interuniversity Attraction Poles initiated by the Belgian Federal
Science Policy Office. The scientific responsibility rests with
its authors. }

\end{document}